\documentclass[twocolumn,showpacs,prl]{revtex4}

\usepackage{graphicx}
\usepackage{amsmath}
\usepackage[center]{subfigure}

\begin{document}
\title{Cascade of Complexity in Evolving Predator-Prey Dynamics}
\author{Nicholas Guttenberg and Nigel Goldenfeld}
\affiliation{Department of Physics and Institute for Genomic Biology, University of Illinois at
Urbana-Champaign, 1110 West Green Street, Urbana, Illinois, 61801-3080.}

\begin{abstract}
We simulate an individual-based model that represents both the
phenotype and genome of digital organisms with predator-prey
interactions.  We show how open-ended growth of complexity arises from
the invariance of genetic evolution operators with respect to changes
in the complexity, and that the dynamics which emerges shows scaling
indicative of a non-equilibrium critical point.  The mechanism is
analogous to the development of the cascade in fluid turbulence.
\end{abstract}


\maketitle

Experiments on digital organisms represent one of the most accurate and
informative methodologies for understanding the process of
evolution\cite{ADAM06}. Systematic studies on digital organisms are
especially informative, because the entire phylogenetic history of a
population can be tracked, something that is much more difficult---but
not impossible\cite{elena2003eem}---to do with natural organisms.
Experiments on digital organisms can be performed over time scales
relevant for evolution, and can capture universal aspects of
evolutionary processes, including those relevant to long-term
adaptation \cite{wilke2002bdo, LENS99}, ecological
specialization\cite{ADAM95, ostrowski2007esa} and the evolution of complex
traits\cite{LENS03}.

Despite this progress, the way in which evolution leads to ever
increasing complexity of organisms remains poorly understood and
difficult to capture in simulations and models to date. Is this because
these calculations are not sufficiently realistic, extensive, or
detailed, or has something fundamental been left out?  In this Letter,
we argue that two fundamental aspects of evolutionary dynamics, with
the character of symmetries, have been omitted, thus causing complexity
growth to saturate.

The first feature is that the evolutionary dynamics must be invariant
with respect to changes in the complexity of the evolving organisms.
That is, if there are inhomogeneities which encourage organisms to have
a specific complexity, then these will act to prevent the complexity of
the system from continually increasing. This invariance is similar in
spirit to that which lies at the heart of the Richardson cascade in
turbulence\cite{richardson1922wpn, KOLM41}. Here, a hierarchy of
length-scales exists due to a transport of energy by scale-invariant
processes between a large length scale and a small length scale.
The largest and smallest features of the flow are determined by where
the invariance is broken. In the biological case, processes invariant
to changes in complexity will allow the dynamics to produce
structures of arbitrarily high complexity.
We will see below, in an explicit model, the effects of different genetic
operations with regard to this invariance criterion. This criterion can
also apply to the way that the fitness of an organism is determined in
the dynamics, either explicitly or implicitly.

The second feature is that there must be some advantage which can only
be gained by an organism in the system being more complex than the
organisms it competes with. Competitive interactions can drive such a
dynamic; for example, if competition can be thought of as one
organism setting the environmental problem that the other organism must
solve. The resulting co-evolution favors an increase in complexity over
a decrease, because for the problem-setter, simplifying the problem
does not exclude an organism already able to solve the problem. This
factor has the same function as viscosity in turbulent flows: it sets
the directionality of the relevant transport.

These two features have precisely the same mathematical role in
evolutionary models as the mechanisms of energy transfer and viscous
dissipation do in fluid turbulence.  Thus, the open-ended growth of
complexity in our model, and the existence of a hierarchy of structures
at all scales in turbulent flows are mathematical consequences of the same
underlying dynamics.  It is not important for this argument what is the
direction of energy flow in the turbulence case: in fact, the direction
depends on dimensionality, with the possibility of the accumulation of
large-scale structures in two-dimensional turbulence through the
so-called inverse cascade.

The implications of this dynamical systems argument are far-reaching,
and impose constraints on how digital evolution models should be built.
For example, despite its popularity, the \lq\lq fitness landscape"
\cite{wright1932rmi, gavrilets2004fla, orr2005gta} picture of evolution
does not satisfy these constraints, and is conceptually insufficient to
account for the open-ended growth of complexity. To illustrate our
points, we now show how open-ended growth of complexity emerges from
underlying dynamical rules in a simple caricature of an evolving
ecosystem.

\smallskip
\noindent {\it Complexity saturation in digital ecosystems:-\/}
Tierra\cite{RAY91} and Avida\cite{ADAM94} are systems of digital
organisms, which are represented as self-replicating programs in a
Turing complete language. In principle any program or behavior can then
be encoded with a sufficiently large genome. In Tierra, organisms exist
in a linear space for which each point in space is associated with an
instruction and replication occurs via a loop which copies the contents
of the space at an offset.  In early work on the Tierra model it became
evident that the dynamics were not neutral with respect to the size of
replicating programs. Evolutionary pressure favored smaller programs as
they replicate with fewer instructions and out-produce the larger
programs in the system. This led to the development of interesting
parasitic behavior in which a program would use a neighbor's
replication code to decrease its length, i.e. the complexity of
organisms did not increase. When this was corrected by a change in the
way in which resources were allotted, the length of organisms was
observed to increase in bursts, but eventually saturated for longer and
longer intervals\cite{RAY91}, a finding attributed to insufficient
richness of the environment\cite{STAN02}.

In Avida, there is a two dimensional grid, each cell of which contains
a program, and replication occurs between cells.  Selection is based on
an organism's ability to solve a particular mathematical problem. Avida
uses an information-theoretic definition of complexity based on the
information learned by the organism from its environment\cite{LENS03}.
For evolution occurring in a single niche, it is found that this
complexity increases for some time, then saturates to a value
determined by the maximum information associated with the niche (the
potential complexity) \cite{ADAM02}.

A similar pattern of saturation in the level of complexity is found in
\lq WebWorld'\cite{CALD98,DROS01,MCKA04,MARO06}.  Here, species are described by a
set of features that may be either present or not, and the total rate
of predation between species is determined by summing over a random interaction matrix
for each feature possessed by the predator and each possessed by the
prey. The total number of features possessed is found to increase in
the presence of interactions above the neutral case. However, the
increase in complexity is eventually limited by the predefined set of
features, there being no possibility of creating new features in the
model.

In summary, these and other digital ecosystems appear to lack the drive
to increasing complexity that arguably is present in real biological
systems.

\smallskip
\noindent {\it Foodchain:-\/} We now present an abstract minimal model
of an evolving predator-prey system, which we call \lq\lq Foodchain".
This model exhibits the potential for an open-ended growth of
complexity. Organisms in this model exist in a two-dimensional space
and interact with each other. The detailed mechanics of replication are
abstracted away (unlike Tierra and Avida)---during replication, genetic
operators (point mutation and gene duplication) are applied to the
genomes, which are of fixed length $2048$, to produce the genome of the
offspring. In \lq Foodchain', fitness is determined solely by
interactions between organisms, as they attempt to eat a random
neighbor each timestep. A certain amount of energy is introduced to
each living organism every time step, and replication occurs when an
organism has an adjacent empty grid cell and a sufficient amount of
energy.

Each organism has a fixed-length string of letters as its genome. These
letters can be upper or lower case, so that each letter is one of 52
possible letters. All but eight letters are inactive and do not
influence the interactions between organisms. Of the eight active
letters, four are offensive (A,B,C,D) and four are defensive (a,b,c,d).

The predator-prey interactions are determined by organisms' genomes. A
particular organism is not predisposed to be predator or prey, and may
even be able to eat its own offspring. The comparison between
genomes consists of matching contiguous substrings of offensive letters
in the organism attempting predation with defensive letters in the
prey. If the predator has a sequence of offensive letters that is not
matched in the prey by a corresponding defensive string, the prey
organism dies and the predator gains a percentage of its energy. A
neutral letter or letter of a different type ends a sequence.

This interaction rule satisfies the condition that fitness in the
system should depend only on relative quantities as well as the
condition that in interactions between different complexities, higher
complexities produce a benefit for the organism. If a particular
organism only has a defensive string of length $L$, then a predator
with an offensive string of length at least $L+1$ will always be able
to eat it; thus there is always a structure at a higher complexity
which can bypass a particular defense.

When an organism replicates, its genome is subject to change from
mutation and other genetic operations. Point mutations occur at a rate
$r_m$ per letter and set the mutated letter to a random letter,
which may be the same as the original. Gene duplication occurs at a rate
$r_d$. In gene duplication, three random values between zero and the
length of the genome are generated: a start position $i_{start}$,
ending position $i_{end}$, and an offset $i_{ofs}$. The sequence
between $i_{start}$ and $i_{end}$ is stored in memory and written back
into the genome starting at $i_{start}+i_{ofs}$. The genome is treated
as being periodic as in microbial DNA, so if $i_{end}<i_{start}$ the
reading process proceeds through the end of the genome and wraps around
to the beginning.

In this system the complexity is taken to be the longest functional
string (separated into attack and defense complexities). The motivation
for this choice is that it is directly related to the capabilities of
the organism. It also represents the interaction between pieces of
information in the organism's genome: together, a sequence of multiple
letters have a certain functionality that, apart, they would not.

Point mutations do not satisfy the condition that the dynamics should
be invariant to changes in complexity. If an organism has a particular
active string of length $L$, there are $L$ chances for a point mutation
to decrease the complexity, and $2$ chances for a point mutation to
increase the complexity. More specifically, if a mutation occurs at the
first letter before or after the string, there is a $1/13$ chance that
the length of the active string increases by $1$. If a mutation occurs
anywhere within the string, there is a $12/13$ chance that the active
length will decrease. The average resultant length $L^\prime$ of an
active string initially of length $L$ after a single point mutation is
given by:
\begin{equation}
\langle L^\prime \rangle = \frac{3}{4}L-\frac{1}{2}-\frac{1}{4L}
\end{equation}
The dynamics of point mutations tends to decrease the active length
because there are many more ways to decrease it than to increase it.
This entropy pressure competes against the selection pressure due to
the advantage that results from having a sequence of higher active
length. The magnitude of the advantage, and thus the selection
pressure, is independent of the absolute sequence length, whereas the
entropy pressure scales with the sequence length. Therefore, there is
an equilibrium active string length (complexity) at which the
entropy pressure is balanced against the selection pressure.

Gene duplication on the other hand operates equally on sequences of
different active lengths so long as the active length is much smaller
than the total genome length. The probability that the gene duplication
region cuts a sequence of length $L$ is $L/L_{genome}$. If a particular
sequence is captured, its length will at least be preserved and may
increase by an amount proportional to the average sequence length in
the organism if the write region is adjacent to another sequence of the
same type.

Point mutations are necessary to fully explore the genetic space, but if
the point mutation rate is too high, the complexity cascade is
inhibited. The next section examines the results of simulations for a
variety of point mutation rates and system sizes in order to probe this
effect.

\begin{figure}[t]
\includegraphics[width=0.9\columnwidth]{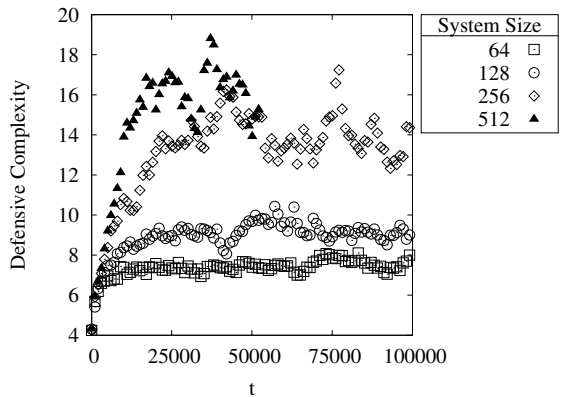}
\caption{Defense complexity versus time in Foodchain for system sizes
64, 128, 256 and 512 square grids. Duplication rate is set to $0.1$ and
mutation rate is set to $0.01$. } \label{FoodchainFig1}
\end{figure}

\begin{figure}[t]
\includegraphics[width=0.9\columnwidth]{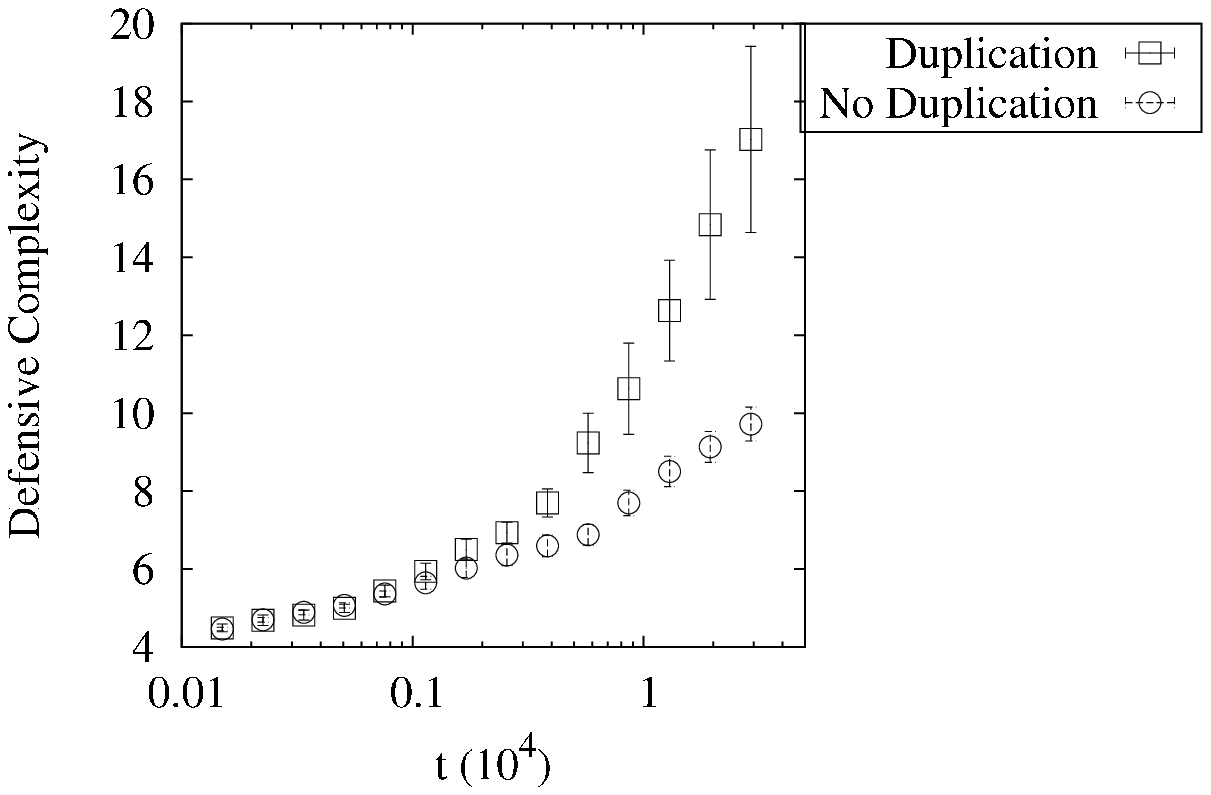}
\caption{Effect of gene duplication on the rate of complexity
increase. Defense complexity is plotted for gene duplication rates of
$0$ and $0.1$. The system is of linear size $256$ and has a mutation rate 
of $0.001$.}
\label{FoodchainFig3}
\end{figure}

Every hundred timesteps the system-wide population, average energy,
average attack complexity, and average defense complexity are stored
for analysis. The attack and defense complexities are taken to be the
longest contiguous string of attack and defense functionality. The
simulation is run for different initial random seeds in order to
extract the mean behavior of these quantities with simulation time.

\smallskip
\noindent {\it Results:-\/} The average defensive complexity of
organisms in the system as a function of time is plotted in Fig.
\ref{FoodchainFig1} for different system sizes. These simulations use a
gene duplication rate (per replication) of $0.1$ and a mutation rate of
$0.01$ per letter. The complexity increases with time for short times,
but then saturates at a value which depends on the system size. We
observed that in a system with no gene duplication, the increase in
complexity was logarithmic with time, whereas the system with gene
duplication exhibited super-logarithmic complexity growth 
(Fig. \ref{FoodchainFig3}). Increasing the system size beyond $256$ has 
diminishing returns, as the change from $256$ to $512$ is less than the change from $128$ to $256$. 

\begin{figure}[t]
\includegraphics[width=0.9\columnwidth]{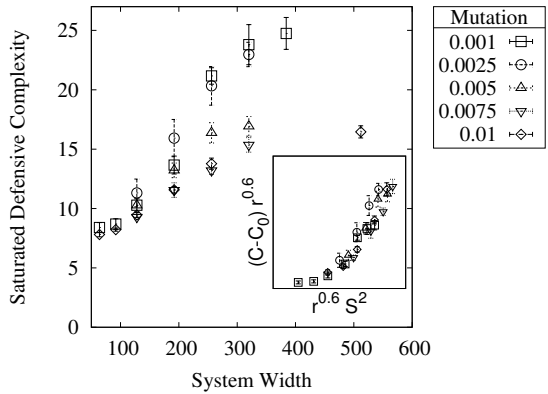}
\caption{Dependence of maximum defensive complexity on system size and
mutation rate. The inset shows that the data collapse onto a single curve
when plotted with a dependent variable $(C-6.65)r^{0.6}$ and
independent variable $r^{0.6} S^2$.}
\label{FoodchainFig2}
\end{figure}

When the mutation rate is decreased to $0.001$, the saturation at low
system sizes is unchanged, but at high system sizes the saturation
point increases. These results are shown in Fig. \ref{FoodchainFig2}.
This suggests that a large mutation rate creates a specific maximum
complexity value due to entropy pressure, and that a small system size
creates a different specific maximum complexity value. Thus the system
will increase in complexity until it reaches the first of those maxima.
When the data are plotted in terms of variables which reflect the
asymptotic complexity scaling, they collapse onto a single curve.
This is analogous to finite size scaling around a critical point in
which the system size creates a departure from criticality and causes
the scaling to saturate. 

The data collapse takes the form of $r^a (C-C_0) = f(r^a S^b)$ where
$f(x)$ scales as $x$ when $x\rightarrow 0$ and $f(x)$ approaches a
constant when $x\rightarrow \infty$. The data are found to collapse for
$a=0.6\pm0.2$, $b=2\pm0.1$, and $C_0 = 6.65\pm0.1$. The error in these
quantities was determined by varying them around the point of best collapse and
monitoring the quality of the collapse. The $S^2$ dependence is indicative
that the total population is the relevant quantity when determining
finite-size effects. The value of $C_0$ is consistent with the
complexity one would generate by randomly generating strings of length
$2048$ with a proportion of defense characters to alphabet size equal
to that observed in the smallest systems. That is to say, at the
asymptote corresponding to high mutation rate and low system size, the
complexity of strings is due entirely to evolutionary pressures on the
relative proportions of the different characters, rather than spatial
organization within the genome.

The saturation due to large point mutation rate can be understood as
being due to its complexity dependence as discussed earlier and in
terms of the Eigen error threshold \cite{EIGE71,WILK05}, but the
observed scaling exponent is not at this time understood. The system
size scaling is surprising as it is not obvious a priori that the
complexity of an organism's genome should be related to the size of the
space the organism lives in (in contrast with turbulence, in which the
complexity of the flow is expressed in the distribution of velocity
throughout the system).

It is possible that the connection between system size and complexity
in \lq Foodchain' is a result of the fixation of complexity-decreasing
mutations. For a finite population of organisms with a set of traits
that may be present or absent in each organism, the fluctuations in the
population and the dynamics of reproduction will eventually cause the
trait to be either present or absent in every member of the population.
The probability of a particular mutation going to fixation is
$P(s)=(1-\exp(-2s))/(1-\exp(-4Ns))$, where $s$ is the selective advantage
and $N$ is the population size\cite{WRIG31,KIMU62,ORR00}.

In the context of the Foodchain model, each organism  may have many
strings of varying complexities only a few of which are responsible for
the organism's reproductive success. The pivotal strings are not
necessarily those of the highest complexity (short defense strings can
still be important in defending against short attack strings held by
other organisms, for instance). However, a mutation to the most complex
string may turn it into a pivotal string even if it is not currently
experiencing selective pressure. In the low mutation rate limit
fixation of complexity-decreasing mutations imposes a limit on the
maximum sustainable length $L$ of a particular string. We balance the
rate of fixation of complexity-increasing mutations (which occur at a
constant rate) with the rate of fixation of complexity-decreasing
mutations (which occur at a rate proportional to $L$): $P(s) - L P(0) =
0$, where $P(0)\propto 1/N$\cite{KIMU62}. This results in the scaling $L \propto N$, consistent with the system size scaling exponent observed in the data collapse.

In the simple \lq Foodchain' model presented here, there is no
separation between primitive organisms that compete with each other
using structures of low complexity and organisms with very complex
offensive and defensive strings.  In order to generate a rich hierarchy
of structures, some form of trophic structure would need to be represented
in the system\cite{drossel2004inf}.

We acknowledge valuable discussions with Carl Woese. Nicholas
Guttenberg was supported by the University of Illinois Distinguished
Fellowship. This material is based upon work supported by the National
Science Foundation under Grant No. NSF-EF-0526747.

\bibliographystyle{apsrev}
\bibliography{complexity}
\end{document}